\documentclass[aps,eqsecnum,nofootinbib,superscriptaddress,showpacs]
{revtex4}
\usepackage[dvips]{graphicx}  
\usepackage{amsmath,amsfonts}
\usepackage{color}  
\def\be{\begin{eqnarray}}
\def\ee{\end{eqnarray}}
\def\beq{\begin{equation}}
\def\eeq{\end{equation}}

\def\p{\partial}

\def\({\left (}
\def\){\right )}
\def\[{\left [}
\def\[{\right ]}

\newcommand{\bra}[1]{\left\langle\, #1\, \right|}
\newcommand{\ket}[1]{\left|\, #1\, \right\rangle}

\begin{document}
\rightline{KEK-TH-1087}
\rightline{hep-th/0605224}

\title{
Extracting information behind the veil of horizon
}

\author{Kengo Maeda}
\email{kmaeda@kobe-kosen.ac.jp}
\affiliation{Department of General Education,
Kobe City College of Technology, Kobe, 651-2194, Japan}

\author{Makoto Natsuume}
\email{makoto.natsuume@kek.jp}
\affiliation{%
Theory Division, Institute of Particle and Nuclear Studies, \\
KEK, High Energy Accelerator Research Organization,
Tsukuba, Ibaraki, 305-0801, Japan}

\author{Takashi Okamura}
\email{okamura@ksc.kwansei.ac.jp}
\affiliation{Department of Physics, Kwansei Gakuin University,
Sanda, 669-1337, Japan}

\date{\today}

\begin{abstract}
In AdS/CFT duality, it is often argued that
information behind the event horizon is encoded
even in boundary correlators.
However, its implication is not fully understood.
We study a simple model which can be analyzed explicitly.
The model is a two-dimensional scalar field propagating
on the s-wave sector of the BTZ black hole
formed by the gravitational collapse of a null dust.
Inside the event horizon,
we placed an artificial timelike singularity
where one-parameter family of boundary conditions is permitted.
We compute two-point correlators with two operators
inserted on the boundary to see
if the parameter can be extracted from the correlators.
In a typical case,
we give an explicit form of the boundary correlators
of an initial vacuum state and show that
the parameter can be read off from them.
This does not immediately imply that
the asymptotic observer can extract the information of the singularity
since one cannot control the initial state in general.
Thus, we also study whether the parameter can be read off
from the correlators for a class of initial states.
\end{abstract}
\pacs{11.25.Tq, 04.20.Dw,04.70.Dy}
\maketitle

\section{Introduction}\label{sec:intro}

It has been conjectured that string theory
on anti-de Sitter~(AdS) spacetime is dual to a gauge theory
or a conformal field theory~(CFT) on the boundary.
This AdS/CFT correspondence often involves black holes.
Then, the singularity problem should be resolved
because nothing is singular in the gauge theory.
However, before we resolve the singularity problem,
one first has to understand how information of the singularity
may be encoded in the boundary correlators.

Maldacena proposed a boundary description
of an eternal AdS-Schwarzschild black hole~\cite{Maldacena01}.
In this description, there are two boundary CFTs
living in each disconnected boundary and
they are coupled to each other via the entangled state.
Then, information behind the horizon should be included
in two-point correlators with each operator inserted on each boundary.
Using the WKB formula, the correlators are approximately obtained
by calculating spacelike geodesics connecting
the two disconnected boundaries~
\cite{LMR00, KOS02,FHKS,excursion,FHMMRS05}.
These geodesics pass through the geometry arbitrarily close
to the spacelike singularity~\cite{FHKS},
so it is argued that some information about the singularity
should be included in the correlators.%

There are several unsolved issues in such a scenario:
\begin{enumerate}
\item[(i)] A black hole is physically formed
by gravitational collapse.
In this case, the black hole has a single exterior
and the boundary theory can have only the correlators
with operators inserted on the single timelike boundary.
It is not clear if information of the singularity
can be extracted from such correlators.
(Such an issue has been actually studied in various contexts.
The AdS/CFT with a single exterior has been studied
in Refs.~\cite{FHKS,FHMMRS05,LM99,BR00,DVK2}.
In particular, Ref.~\cite{FHMMRS05} argues that
information behind the horizon is reflected
into subleading behaviors of the correlators).

\item[(ii)] Most works use the WKB approximation.
The WKB technique is useful to see the underlying physics,
but the analytic expression is certainly desirable.

\item[(iii)] These works pinpoint {\it where}
information of singularities is encoded in the correlators
(namely, it appears in the light-cone singularities
of the correlators with a specific weight due to the null geodesics
reaching spacelike singularities),
but {\it how} information of singularities is encoded
is not understood yet.

\item[(iv)] The notion of ``detectability" is not very clear.
As we will see, the precise form of the boundary correlators
depends on the details of the singularity,
but this does not immediately imply that
the asymptotic observer can reconstruct information of the singularity.
These two are entirely different notions,
but they are not clearly distinguished in the literature.
\end{enumerate}

This paper addresses these issues by studying a simple model.
We investigate how information about singularity
inside a single exterior AdS black hole
is encoded in the bulk correlators outside the horizon [issue~(i)].
We consider the BTZ black hole formed by the gravitational collapse
of a null dust.
In order to explicitly represent ``information" of singularity,
an artificial timelike singularity is placed inside the black hole
by removing a single spatial point [issue~(iii)].
The information is reflected simply in the boundary condition
at the singularity and, as we will see,
the boundary condition can be represented by real one parameter.
Namely, this parameter represents the information of our singularity.
In principle, the boundary condition should be determined
by the full theory of quantum gravity,
but it is not our purpose here to determine the boundary condition.
Instead our interest is to see how the parameter appears
in the correlators outside the horizon.%
\footnote{This spacetime also has a usual spacelike singularity
(see Fig.~\ref{test}), but we do not address the issue
how the information of the spacelike singularity is encoded
in the correlators.
This question may be addressed along the same line as previous works~
\cite{LMR00, KOS02,FHKS,excursion,FHMMRS05}.}

We consider a conformal scalar field propagating on the black hole.
This model can be fully analyzed [issue~(ii)].
Since the whole evolution is determined
by both the boundary condition and the initial state,
the correlator with operators inserted
on the outer-communication region is indirectly affected
also by the boundary condition behind the horizon.
We show that the parameter can be restored
from the asymptotic behavior of the boundary correlator
when the initial state is a vacuum state.
So, the boundary condition behind the horizon is detectable,
if the vacuum state is uniquely selected as the initial state.
We discuss this possibility in the final section.

In the case that one cannot control the initial state,
the asymptotic observer cannot necessarily extract
the information of the singularity,
as the observer has access to only part of the Cauchy surface.
Suppose that the correlator for the vacuum state
at a specific real parameter cannot be reproduced
by the correlators for any excited states at another real parameter.
Then, it is reasonable to say that
the observer can extract the information even without knowing
the initial state.
Thus, we also calculate the correlators for a class of excited states
[issue~(iv)].
Although our result is not conclusive at this point,
we show that the correlator for the vacuum state
at a specific parameter cannot be reproduced by the correlators
for a class of excited states we consider.

In this paper, we do not consider the precise correspondence
with the boundary theory, {\it i.e.},
we do not know the boundary theory,
and we do not derive the boundary physics from the boundary theory.
Instead, our purpose in this paper is
how the would-be boundary correlators may contain
the information of the singularity.
But we give some general remarks in Sec.~\ref{sec:conclusion}.

The plan of our paper is as follows.
In the next section, we construct the model of black hole
formed by the gravitational collapse of a null dust.
In Sections~\ref{sec:Wightman_construction} and \ref{sec:results},
we explicitly calculate Wightman function of a conformal scalar field
for a vacuum state%
\footnote{As shown in Ref.~\cite{Book},
the Wightman function can be obtained
by Fourier transform of the detector response function,
which is constructed by counting particles from the black hole.
}
and see if the parameter is reflected in the functions.
In Sec.~\ref{sec:state_dep}, we also calculate the function
for some excited states and argue the possibility
to derive the information from generic initial states.
Conclusion and discussion are devoted in the final section.

\section{Preliminaries}\label{sec:toy_model}

We consider an AdS black hole
with a single exterior boundary.
Such a black hole is usually formed by gravitational collapse
and the spacetime inside the horizon ends
at a spacelike~(or possibly null) singularity.
Information about such a singularity would be obtained
from the initial data by solving the dynamics in the past direction.
In general, it would be very complicated
due to the time-dependence of the background spacetime
and the non-linearity of dynamics near the singularity.
On the other hand, for a timelike singularity,
one can reflect some of its properties simply
as the boundary conditions.
So, information about a timelike singularity is quantitatively
more tractable than the one about a spacelike singularity.
In this section, we construct an artificial model
of a timelike singularity characterized by one-parameter family
of boundary conditions.

\subsection{Model spacetime}\label{sec:model_spacetime}

Our model spacetime is two-dimensional spacetime
whose metric is given by
\beq
ds^2 = - f(r)\, dv^2 + 2\,dv\,dr
     = - f(r)\, dt^2 + \frac{dr^2}{f(r)} \qquad (r > 0),
\label{eq:model_metric}
\eeq
where $dt := dv - dr/f(r)$ and
\begin{align}
   f(r)
  &= \begin{cases}
           (r-r_0) (r+r_0) & \text{~~~for~~} v \ge 0 \\
                     1+r^2 & \text{~~~for~~} v< 0 \end{cases}~.
\label{eq:def-metric_f}
\end{align}
The spacetime has the event horizon;
for $v \ge 0$, it is located at $r=r_0$.
The surface gravity $\kappa$ for the spacetime
is given by $\kappa := f'(r_0)/2 = r_0$.

The metric is motivated by the BTZ black hole formed
by the gravitational collapse of a null dust shell \cite{Husain}.
The three-dimensional metric
\begin{align}
  & ds^2 = - f(v,r)\, dv^2 + 2\, dv\, dr + r^2\, d\theta^2~,
\label{eq:assumption-metric}
\end{align}
is a solution of the Einstein equations
with cosmological constant $\Lambda$ and
with the energy-momentum tensor
of the cylindrically symmetric null dust
\begin{align}
  & T_{\mu\nu} = \frac{\rho(v)}{4\pi r}\, \p_\mu v\, \p_\nu v~,
\end{align}
provided that
\begin{align}
  & f(v,r) = - \Lambda\, r^2 - g(v)~,
& & \rho(v) = d g(v)/dv~.
\end{align}

\begin{figure*}[htb]
  \begin{center}
    \includegraphics[clip]{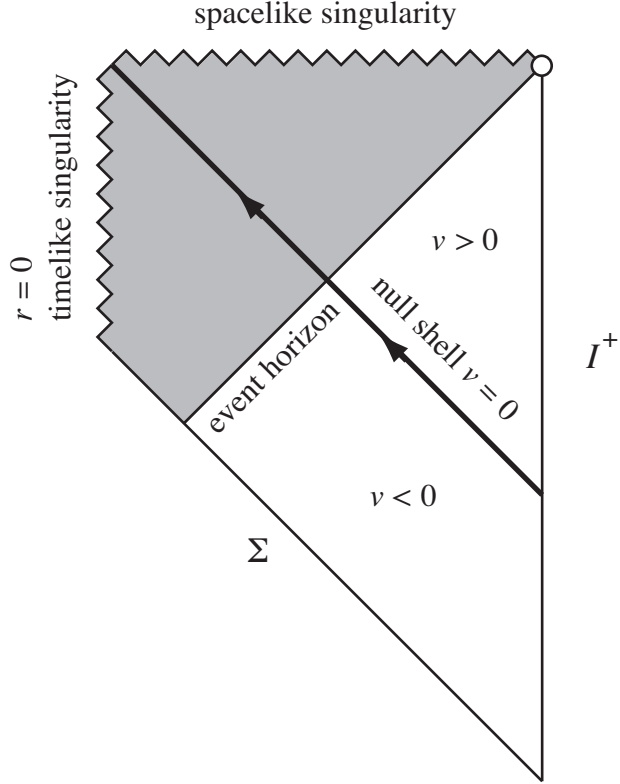}
  \end{center}
\caption{\label{test}
Penrose diagram of the BTZ black hole
formed by the gravitational collapse of a null dust shell.
The spacetime for $v<0$ corresponds to $\text{AdS}_3$
with a timelike singularity located at $r=0$.
The surface $\Sigma$ is an initial Cauchy surface.
}
\end{figure*}

Consider a thin shell made of the null dust falling
along the $v=0$ surface.
Then, the energy density is represented by $\rho(v)=\rho_0\delta(v)$.
If $\rho_0$ is large enough,
the BTZ black hole is formed after the collapse of the shell
(see Fig.~\ref{test}).
We set the cosmological constant $\Lambda$ to $-1$
and we use $\text{AdS}_3$ ($g=-1$) for the metric
before the collapse.%
\footnote{Strictly speaking, a timelike singularity is
artificially added simply by removing a single spatial point
from $\text{AdS}_3$ as described in Sec.~\ref{sec:model_field}.}
Then, the two-dimensional $(v, r)$-section
of Eq.~(\ref{eq:assumption-metric}) is just our model metric
(\ref{eq:model_metric}).%
\footnote{The horizon radius $r_0$ is related
to the energy density of the shell by $r_0=\sqrt{\rho_0-1}$.
}
%
%
%

From the two-dimensional point of view,
the metric~(\ref{eq:model_metric}) is obtained from
the two-dimensional Jackiw-Teitelboim gravity %
\cite{ref:Liouville},
\begin{align}
  S = \int dvdr~\sqrt{-g}~e^\psi ( R - 2 \Lambda )~,
\label{eq:two_dim-action}
\end{align}
where $g_{\mu\nu}$ and $R$ are the two-dimensional metric
and the scalar curvature, respectively.
For the solution (\ref{eq:model_metric}),
the field $\psi$ is given by $e^\psi=r$
and is continuous across the null dust shell.

\subsection{Model field}\label{sec:model_field}

We consider the evolution
of a massless test scalar field $\phi$ which is minimally coupled
to the Jackiw-Teitelboim gravity:
\begin{align}
  S = \int dvdr~\sqrt{-g}~\left[~e^\psi ( R - 2 \Lambda )
    - \frac{1}{2}\, (\nabla\phi)^2~\right]~.
\label{eq:two_dim-action_scalar}
\end{align}
Its field equation becomes
\be
\label{two-scalar-eq}
(-\p_t^2+\p_{r_*}^2) \phi=0,
\ee
where $t:=v-r_*$ and
\begin{align}
  & r_* := \int \frac{dr}{f(r)}
  = \begin{cases}
      \dfrac{1}{2 \kappa}\, \ln \dfrac{r-r_0}{r+r_0}
           & \text{~~~for~~} v \ge 0 \\ \\
      \tan^{-1}(r) & \text{~~~for~~} v < 0
    \end{cases}~.
\end{align}

Note that $v$ and $r$ are global coordinates,
but $r_*$ and $t$ are local ones.
In order to distinguish between the local coordinates
in the region $v \ge 0$ and those in the region $v < 0$,
we often use the superscript ${}^+$ (${}^-$)
representing the local coordinates
in the region $v \ge 0$ ($v < 0$).
Then, the retarded coordinates $u^+ := t^+ - r_*^+$
and $u^- := t^- - r_*^-$ are related to each other
on the surface $v=0$ as
\begin{align}
  & e^{- \kappa u^+}
  = \left\vert\, \frac{\tan (u^-/2) + r_0}{\tan (u^-/2) - r_0}\,
    \right\vert
& & (- \pi < u^- \le 0)~.
\label{eq:relation-u}
\end{align}

Let us remove a single spatial point $r=0$
from the spacetime to make an artificial model
of a timelike singularity;
this model can be fully analyzed~(see Fig.~\ref{test}).
Since we remove the single spatial point $r=0$,
we should specify a boundary condition for the scalar field there.
To quantize the scalar field,
the conserved Klein-Gordon (KG) inner product is necessary,
so we require that the KG inner product is conserved.
From the equation of motion of the scalar field,
the variation of the inner product is given by
\begin{align}
   (\phi,\,\psi)_{\Sigma_1} - (\phi,\,\psi)_{\Sigma_0}
  &= i \int^{\Sigma_1 \cap (r=\infty)}_{\Sigma_0 \cap (r=\infty)}
     d\Sigma^\mu~\big( \phi^* \nabla_\mu \psi
      - \psi \nabla_\mu \phi^* \big)\big\vert_{r=\infty}
\nonumber \\
  &- i \int^{\Sigma_1 \cap (r=0)}_{\Sigma_0 \cap (r=0)}
     d\Sigma^\mu~\big( \phi^* \nabla_\mu \psi
      - \psi \nabla_\mu \phi^* \big)\big\vert_{r=0}~,
\end{align}
where $\Sigma_0$ and $\Sigma_1$ are spacelike surfaces.

We impose the Dirichlet boundary condition $\phi(r=\infty) = 0$
at null infinity to mimic the fall-off condition of the scalar field
in higher-dimensional asymptotically AdS spacetime.
Then, the solution is represented
by two arbitrary functions, $A^+$ and $A^-$ as
\begin{align}
  & \phi(t,r_*)
  = \begin{cases}
      A^+(u^+) - A^+(v) & \text{~~~for~~} v \ge 0 \\
      A^-(u^-) - A^-(v - \pi) & \text{~~~for~~} v < 0
    \end{cases}~.
\label{eq:general-sol}
\end{align}

Then, for any fields $\phi$ and $\psi$
satisfying Eq.~(\ref{two-scalar-eq}),
the conservation of the KG inner product implies
\beq
   0 = \big( \phi^*\, \partial_{r_*} \psi
      - \psi\, \partial_{r_*} \phi^* \big)\big\vert_{r=0}~.
\eeq
This condition is equivalent to
\begin{align}
  & \left. \frac{\p_{r_*}\phi}{\phi} \right\vert_{r=0} = c~,
\label{eq:bc}
\end{align}
where $c$ is a real number which is common to all $\phi$.%
\footnote{
This real one-parameter family of the boundary conditions
corresponds to the self-adjoint extensions
of the time translation operator,
$\partial_{r_*}^2$ \cite{ref:ReedSimon}. }
Thus, we have a model of scalar field whose evolution
is characterized by the parameter $c$
via the boundary condition at the timelike singularity.
The conditions $c=0$ and $c=\infty$ correspond to
Neumann and Dirichlet boundary conditions, respectively.

This spacetime also has a usual spacelike singularity
due to the gravitational collapse (see Fig.~\ref{test}),
but we do not address the issue
how the information of the spacelike singularity is encoded
in the correlators.

We choose the initial Cauchy surface $\Sigma$
so that it intersects with the horizon.
This condition is necessary in order for the timelike singularity
not to be visible to the asymptotic observer.
If one chooses a null surface as $\Sigma$,
this requirement is always met by placing $\Sigma$
arbitrarily close to the $v=0$ surface.

Under the WKB approximation~\cite{LMR00,excursion},
a two-point correlator has been obtained
by calculating the geodesic distance between two points.
This approximation is based on particle picture
propagating along the geodesics.
In our case, the particle corresponds to a massless particle
propagating along the null geodesics between two points.
In field theory picture, the parameter represents
how the ``wave packet'' $\phi$ is reflected on the boundary, namely,
it determines the phase shift.
But in particle picture, it is not clear
how the parameter $c$ appears in the two-point correlator.
In the next section,
we explicitly calculate a two-point correlator
with two operators inserted on the timelike boundary
at null infinity without using WKB formula.

\section{Construction of the bulk Wightman function}
\label{sec:Wightman_construction}

Our aim is to see if the information of the singularity appears
in the two-point correlator by explicitly constructing
the Wightman function.
The Wightman function is defined by
\begin{align}
  & W_\Psi( x_1, x_2\, ; c )
  := \bra{\Psi\,;c}\, \phi(x_1)\, \phi(x_2)\, \ket{\Psi\,;c}~,
\label{def-wight}
\end{align}
where $x_1$ and $x_2$ are two spacetime points and
$\ket{\Psi\,;c}$ is an initial quantum state on $\Sigma$
with boundary condition specified by $c$.
There are many choices for the initial state $\ket{\Psi\,;c}$,
so we need to see the $c$-dependence for all possible states
to extract the information of the singularity.
This state dependence is discussed in Sec.~\ref{sec:state_dep}.
In this section,
we pay attention to a vacuum state as the initial state.

\subsection{Mode function}\label{sec:mode_func}

First, we explicitly construct the mode solution
which satisfies the Dirichlet condition at null infinity
and Eq.~(\ref{eq:bc}) at the origin in the $v<0$ region.
Then, we extend the mode solution into the $v\ge0$ region
by the continuity condition across the $v=0$ null surface.

In the AdS region $v<0$, a positive frequency mode $f_n$ is given by
\begin{align}
   f_n(t,r_*)
  &= \sqrt{ \frac{2}{ \pi\omega_n-\sin\pi\omega_n } }~
     e^{-i\omega_n t^-} \sin\big[~\omega_n (r_*^- - \pi/2 )~\big]~.
\label{eq:mode-f_AdS}
\end{align}
The mode solution is a normalized solution of the form
(\ref{eq:general-sol}) satisfying the boundary condition (\ref{eq:bc}).
The KG inner product is used for normalization.
The frequencies $\omega_n$ must satisfy the eigenvalue equation%
\footnote{
When $c \ge 0$, there are only real solutions for Eq.~(\ref{omega-eq}).
Hereafter, we consider this case.
}
\begin{align}
  & 0
  = \omega_n + c\, \tan\left(\frac{\pi\omega_n}{2}\right)
& & ( \omega_n > 0 )~.
\label{omega-eq}
\end{align}
The frequencies $\omega_n$ are ordered
such that $\omega_n>\omega_m$ for $n>m$.
Comparison of Eq.~(\ref{eq:mode-f_AdS}) with Eq.~(\ref{eq:general-sol})
leads to
\begin{align}
   A^-(u^-)
  &= \frac{ e^{-i\omega_n (u^- + \pi/2)} }
          { i\, \sqrt{2 ( \pi\omega_n-\sin\pi\omega_n )} }~.
\label{eq:R_minus}
\end{align}

From the continuity of the mode function,
one can extend the mode function $f_n$ obtained in the $v < 0$ region
into the $v \ge 0$ region.
The continuity condition on the null surface $v=0$ becomes
\begin{align}
  & A^+(u^+) - A^+(0)
  = A^-(u^-) - A^-(-\pi)~.
\label{eq:continuityI}
\end{align}
The above equation (\ref{eq:continuityI}) implies
\begin{align}
  & A^+(u^+) = A^-\big( u^-(u^+) \big)
  = \frac{ \exp\left[ -i\omega_n \big\{~u^-(u^+) + \pi/2~\big\}\,
               \right] }
         { i\, \sqrt{2 ( \pi\omega_n-\sin\pi\omega_n )} }~,
\label{eq:continuityII}
\end{align}
where we use the relation $u^+ \vert_{u^-=-\pi} = 0$
obtained from Eq.~(\ref{eq:relation-u}).

We are especially concerned with
the behavior of the Wightman function outside the horizon.
Outside the horizon,
the function $u^-(u^+)$ is given by
\begin{align}
  & u^-(u^+)
  = - 2~\tan^{-1}\left( r_0\,
     \frac{ 1 + e^{- \kappa u^+} }{ 1 - e^{- \kappa u^+} } \right)
  =: -2\, h(u^+)~,
\label{eq:relation-uII}
\end{align}
and~$\tan^{-1}(r_0) <  h(u^+) < \pi/2$.
Thus, the mode function in the $v \ge 0$ region is written as
\begin{align}
   f_n(t, r_*)
  &= \frac{ e^{- i \pi \omega_n/2} }
          { i\, \sqrt{2 ( \pi\omega_n-\sin\pi\omega_n )} }~
     \left( e^{ 2 i \omega_n h(u^+) }
          - e^{ 2 i \omega_n h(v) } \right)~.
\label{eq:mode-f_BH}
\end{align}
%

\subsection{Wightman function}\label{sec:Wightman}

One can expand the scalar field $\phi(x)$ by the mode function $f_n$ as
\be
\label{expansion}
\phi(x) = \sum_{n=1}^\infty~
  \left( a_n\, f_n(x) +a_n^\dagger\, f_n^*(x) \right),
\ee
where $a_n$ and $a_n^\dagger$ are the annihilation
and creation operators, respectively.
Then, $a_n|0;c>=0$ defines the vacuum state associated with $f_n$
which satisfies the boundary condition (\ref{eq:bc}).

The Wightman function for the vacuum state is given by
\begin{align}
  & W_0( x_1\, ; x_2\, ;c )
  = \bra{0\,;c}\, \phi(x_1)\, \phi(x_2)\, \ket{0\,;c}~.
\end{align}
In the black hole region $v,\,v'>0$,
the Wightman function is written by
\begin{align}
   W_0( u^+_1, v_1\, ; u^+_2, v_2\, ; c )
  &= \mathcal{W}\big(\, h(u^+_1) - h(u^+_2)\, \big)
  - \mathcal{W}\big(\, h(u^+_1) - h(v_2)\, \big)
\nonumber \\
  &- \mathcal{W}\big(\, h(v_1) - h(u^+_2)\, \big)
  + \mathcal{W}\big(\, h(v_1) - h(v_2)\, \big)~,
\label{eq:Wightman_late}
\end{align}
where
\begin{align}
  & \mathcal{W}(\alpha)
  := \frac{1}{2}\, \sum_{n=1}^\infty~
     \frac{ e^{2 i \omega_n\, \alpha} }
          { \pi\omega_n - \sin(\pi\omega_n) }
& & ( |\alpha| < \pi/2)~.
\label{eq:def-calW}
\end{align}
As a check, Eq.~(\ref{eq:Wightman_late}) consists of two contributions
(terms with plus sign and those with minus sign).
This is characteristic to the Dirichlet boundary condition
at $u^+ = v$. (Recall the method of images in electrostatics.)

The infinite sum in Eq.~(\ref{eq:def-calW}) can be written
in terms of an integration (See Appendix~\ref{sec:appendix}):
\begin{align}
   \mathcal{W}(\alpha)
  &= \frac{c}{2 \pi}\,
    \int^\infty_0 \frac{dy}{y}~
     \frac{ \cosh(2\alpha y) - 1 }
          { y \big( 1 + \cosh(\pi y) \big) + c \sinh(\pi y) }
  + \frac{1}{4\pi}\, \ln\left\vert\, \cot\alpha\, \right\vert
\nonumber \\
  &- \frac{i}{2}\, \frac{c}{2 + \pi c}\, \alpha
  + \frac{i}{8}\, \text{sgn}(\alpha)~.
\label{eq:calW}
\end{align}
Therefore, the bulk Wightman function is
\begin{align}
  & W_0( u^+_1, v_1\, ; u^+_2, v_2\, ; c )
  = \frac{c}{2 \pi}\, \int^\infty_0 \frac{dy}{y}~
       \frac{ Q(y\, ; u^+_1, v_1\, ; u^+_2, v_2\, ; c ) }
            { y\, \big( 1 + \cosh(\pi y) \big) + c \sinh(\pi y) }
\nonumber \\
  &+ \frac{1}{4\pi}\, \ln\left\vert\,
     \frac{ \cot\big[\, \Delta(u^+_1, u^+_2)\, \big]
      \cdot \cot\big[\, \Delta(v_1, v_2)\, \big] }
          { \cot\big[\, \Delta(u^+_1, v_2)\, \big]
      \cdot \cot\big[\, \Delta(v_1, u^+_2)\, \big] }\, \right\vert
  + \frac{i}{8}\, \Big(
    \text{sgn}\big[\, \Delta(u^+_1, u^+_2)\, \big]
\nonumber \\
  &\hspace*{2.0cm}
  + \text{sgn}\big[\, \Delta(v_1, v_2)\, \big]
  - \text{sgn}\big[\, \Delta(u^+_1, v_2)\, \big]
  - \text{sgn}\big[\, \Delta(v_1, u^+_2)\, \big] \Big)~,
\label{eq:bulk-Wightman_late}
\end{align}
where $\Delta(s_1, s_2) := h(s_1) - h(s_2)$ and
\begin{align}
  & Q(y\, ; u^+_1, v_1\, ; u^+_2, v_2\, ; c )
  := \cosh\big[\, 2 \Delta(u^+_1, u^+_2)\, y\, \big]
          + \cosh\big[\, 2 \Delta(v_1, v_2)\, y\, \big]
\nonumber \\
  &\hspace*{4.0cm}
  - \cosh\big[\, 2 \Delta(u^+_1, v_2)\, y\, \big]
  - \cosh\big[\, 2 \Delta(v_1, u^+_2)\, y\, \big]
\nonumber \\
  =& - 4 \sinh\big[\, \Delta(u^+_1, v_1)\, y\, \big]
         \sinh\big[\, \Delta(u^+_2, v_2)\, y\, \big]
\nonumber \\
  &\hspace*{3.0cm}
  \times \cosh\Big[ \big(\, \Delta(u^+_1, v_2)
                      + \Delta(v_1, u^+_2)\, \big)\, y\, \Big]~.
\label{eq:def-Q}
\end{align}
Note that the Wightman function outside the horizon
(\ref{eq:bulk-Wightman_late})
depends on $c$ only through the real part and that the imaginary part,
which is the expectation value of commutator of the scalar field,
is independent of $c$.
Related issues are discussed in the next section.

It is instructive to see the behavior of the function~
(\ref{eq:Wightman_late}) near the infinity $r_1 \rightarrow \infty$
and at the late time $t_1 \rightarrow \infty$:
\begin{align}
  & W_0( u^+_1, v_1\, ; u^+_2, v_2\, ; c )
\nonumber \\
  \sim& \frac{2}{r_1}\, h'(t^+_1)~\left[~
     \mathcal{W}\big(\, h(t^+_1) - h(u^+_2)\, \big)
  - \mathcal{W}\big(\, h(t^+_1) - h(v_2)\, \big)~\right]
\nonumber \\
  \sim& - \frac{4 \kappa r_0}{1 + r_0^2}\,
    \frac{ e^{- \kappa t^+_1} }{r_1}\,
  \left[~\mathcal{W}\big(\,  \tan^{-1}(r_0) - h(u^+_2)\, \big)
  - \mathcal{W}\big(\,  \tan^{-1}(r_0) - h(v_2)\, \big)~\right]~
\label{eq:Wightman_remote_late}
\end{align}
under the approximation
\begin{align}
  & u^+ = t^+ - r^+_* \sim t^+ + \frac{1}{r}~,
& & v = t^+ + r^+_* \sim t^+ - \frac{1}{r}~.
\end{align}
In the third equation, the large $u^+$ expansion for $h(u^+)$
is used [See Eq.~(\ref{eq:relation-uII})]:
\begin{align}
   h(u^+)
  = \tan^{-1}(r_0) + \frac{2 r_0}{1 + r_0^2}\, e^{- \kappa u^+}
  + O\big( e^{- 2 \kappa u^+} \big)~.
\label{eq:late-h}
\end{align}
The scalar field decays exponentially,
so the field loses the late-time correlation;
this is due to the redshift factor and
is common to black hole spacetimes.
The quantum fields in a black hole background is expected
to become a thermal state at temperature $T_{\rm BH}=\kappa/(2\pi)$
due to the Hawking radiation and that
the time scale of its thermalization
is the order of magnitude of $O(1/T_{\rm BH})$.
The form of Eq.~(\ref{eq:Wightman_remote_late})
is consistent with the expectation.

\section{Boundary Wightman function}\label{sec:results}

According to the standard AdS/CFT procedure~\cite{BDHM,Kle-Wit},
the boundary Wightman function $G_\Psi( t^+_1, t^+_2\, ; c )$
is obtained by
\begin{align}
  & G_\Psi( t^+_1, t^+_2\, ; c )
  = \lim_{r_1,r_2 \rightarrow \infty}~
     r_1\, r_2~W_\Psi( t^+_1, r_1\, ; t^+_2, r_2\, ; c )~.
\label{eq:AdS_CFT}
\end{align}
Then, the Wightman function $W_0$ behaves near the infinity as
\begin{align}
  & W_0( u^+_1, v_1\, ; u^+_2, v_2\, ; c )
  \sim - \frac{ 4\, h'(t^+_1)\, h'(t^+_2) }{r_1\, r_2}~
      \mathcal{W}''\big[ \Delta(t^+_1, t^+_2) \big]~.
\label{eq:boundary-Wightman_late-pre}
\end{align}
Thus, the boundary Wightman function
$G_0( t^+_1, t^+_2\, ; c )$ for the vacuum becomes
\begin{align}
   G_0( t^+_1, t^+_2\, ; c )
  &= - \frac{ 4\, h'(t^+_1)\, h'(t^+_2) }{\pi}~
  \left(~2 c\, \int^{\infty}_0 dy~
     \frac{ y\, \cosh\big[~2 \Delta(t^+_1, t^+_2)\, y~\big] }
          { y \big( 1 + \cosh(\pi y) \big) + c \sinh(\pi y) }
  \right.
\nonumber \\
  &\left. \hspace*{3.0cm}
  + \frac{ \cos\big[ 2 \Delta(t^+_1, t^+_2) \big] }
         { \sin^2\big[ 2 \Delta(t^+_1, t^+_2) \big] }
  + \frac{i \pi}{4}\, \delta'\big[\, \Delta(t^+_1, t^+_2)\, \big]~
  \right)~.
\label{eq:boundary-Wightman_late}
\end{align}
The imaginary part is independent of $c$,
which reflects the state-independence of the imaginary part
of the bulk Wightman function.

\begin{figure*}[htb]
  \begin{center}
    \includegraphics[width=7.0cm,height=5.5cm,clip]{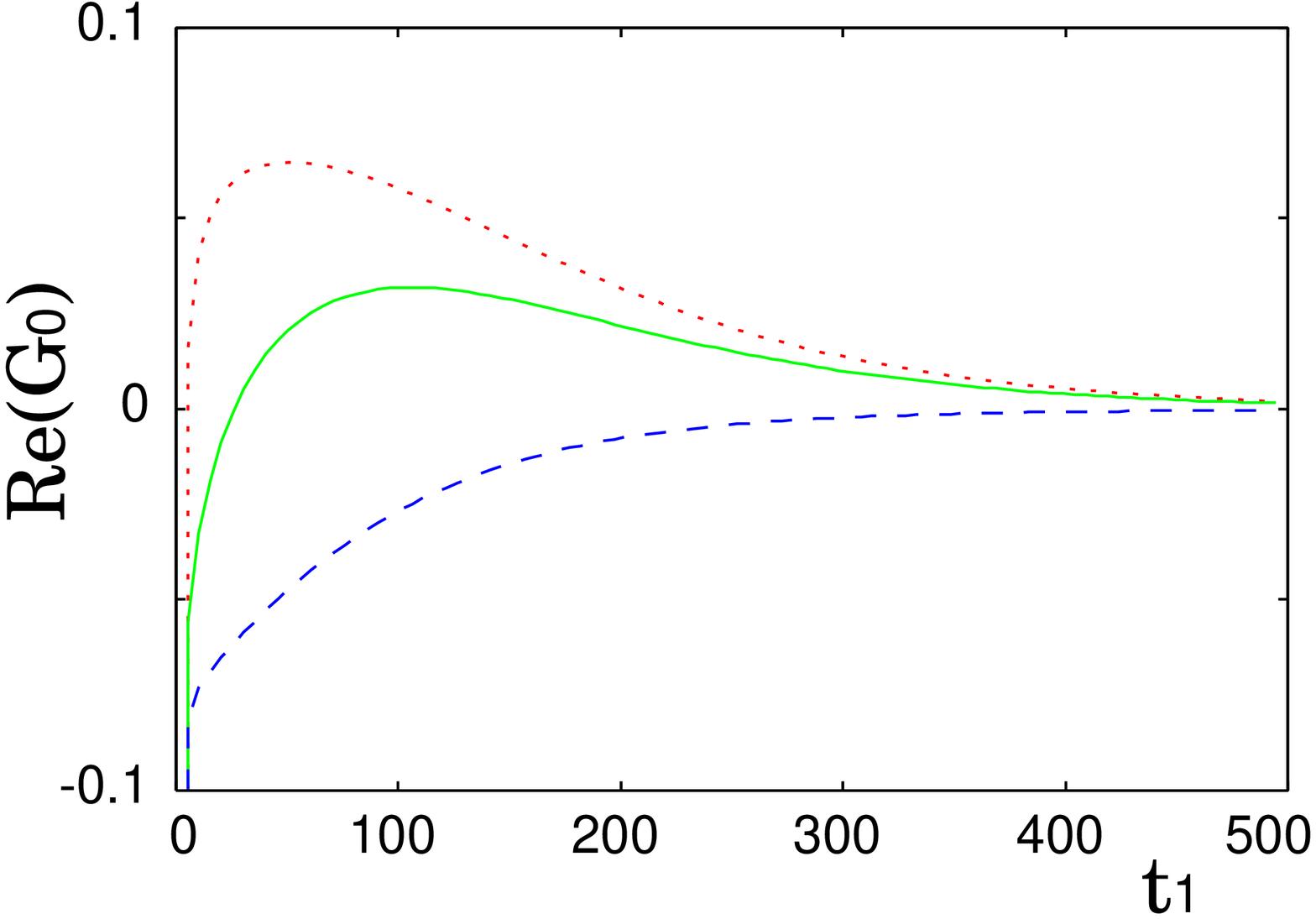}
\hfill
    \includegraphics[width=7.0cm,height=5.5cm,clip]{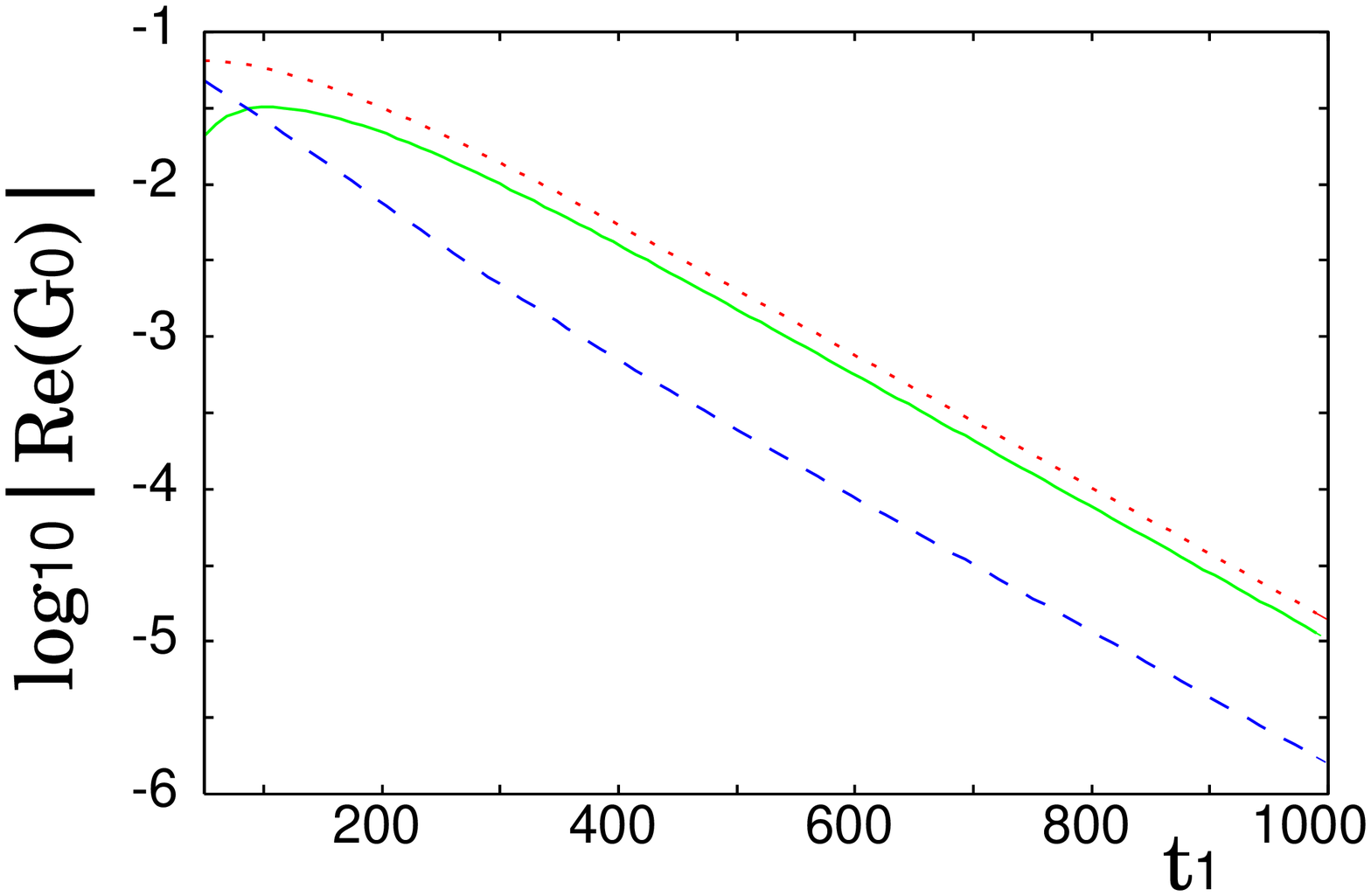}
  \end{center}
\caption{\label{fig3}
(color online)
The real part of the two-point function
$G_0(t_1, 10^{-5}\, ;\, c )$.
The dotted, solid, and dashed lines correspond to
$c=1$, $c=10$, and $c=100$, respectively.}
\end{figure*}

Let us see some specific examples.
For $c=0$ and $c=\infty$, the boundary Wightman function
$G_0( t^+_1, t^+_2\, ; c )$ can be written in analytic form:
\begin{align}
  & \text{Re} \left[~G_0( t^+_1, t^+_2\, ; c=0 )~\right]
  = - \frac{ 4\, h'(t^+_1)\, h'(t^+_2) }{\pi}~
    \frac{ \cos\big[ 2 \Delta(t^+_1, t^+_2) \big] }
              { \sin^2\big[ 2 \Delta(t^+_1, t^+_2) \big] }~,
\label{boundary-wight1} \\
  & \text{Re} \left[~G_0( t^+_1, t^+_2\, ; c=\infty)~\right]
  = - \frac{ 4\, h'(t^+_1)\, h'(t^+_2) }{\pi}~
    \frac{1}{ \sin^2\big[ 2 \Delta(t^+_1, t^+_2) \big] }~.
\label{boundary-wight2}
\end{align}
We also plot the real part of $G_0$ in Fig.~\ref{fig3}
for several values of $c$.

%
%

As is clear from Eq.~(\ref{eq:boundary-Wightman_late}),
the information behind the horizon
(the boundary condition set behind the horizon) is imprinted on
the boundary Wightman function in the outer communicating region.
This fact is not due to the causality violation
but due to the entanglement between two states inside
and outside the horizon on the initial Cauchy surface $\Sigma$.
So, it is generic to the models with
the timelike singularity such as ours.
To see this,
let us consider the evolution of the Wightman function
for general black hole spacetime.

From the field equation of $\phi$, 
the Wightman function for a linear scalar field
in the outer communicating region evolves as
\begin{align}
  & W_\Psi( x_1\, ; x_2\, ; c )
\nonumber \\
  =& \int_{\Sigma} d\Sigma^\mu_{y_1}\,
    \int_{\Sigma} d\Sigma^\nu_{y_2}~
    G_R( x_1\, ; y_1 )\, G_R( x_2\, ; y_2 )
    \stackrel{\leftrightarrow}{\nabla}{}^{y_1}_\mu
    \stackrel{\leftrightarrow}{\nabla}{}^{y_2}_\nu~
    W_\Psi( y_1\, ; y_2\, ; c )~,
\label{eq:develop-Wightman}
\end{align}
where $x_1$ and $x_2$ are in the outer communicating region
and $\Sigma$ is an initial Cauchy surface.
The function $G_R(x; y)$
is the retarded Green function of the equation of motion
for the scalar field,
so it is causal and independent of a quantum state.
So, it is independent of $c$ in the outer-communicating region.
Therefore,
the $c$-dependence on the Wightman function could come
only from the initial Cauchy surface $\Sigma$
in the outer communicating region, and it indeed comes as follows.

Since the event horizon is just a null surface
dividing the spacetime into two regions,
it is unnatural to set the quantum state $\Psi$
by a product of states inside and outside the horizon.
Physical quantities constructed from the energy-momentum tensor
could diverge on the horizon.%
\footnote{For example,
let us consider the Rindler vacuum in the Minkowski spacetime.
It is well-known that
the Rindler vacuum has a product form of the states
in the right wedge and in the left wedge,
and its energy-momentum tensor diverges on the Rindler horizon.}
So, the initial state in the outer communicating region
could be entangled with the one behind the horizon.
This implies that the information about the boundary condition
is imprinted on the boundary Wightman function
through the entanglement on $\Sigma$ for general $\Psi$.

The initial value of the bulk Wightman function
depends not only on the boundary condition $c$,
but also on the choice of the quantum states.
Therefore, the detectability of the information behind the horizon
is a delicate problem,
though the information behind the horizon
is imprinted on the boundary Wightman function in principle.

\section{Wightman function for squeezed states}\label{sec:state_dep}

In this section,
we discuss whether we can derive the information
about parameter $c$ from $G_\Psi( t^+_1, t^+_2\, ; c )$
when the initial state is not restricted to the vacuum state.
The reason why such a consideration is necessary
is because the asymptotic observer has no way
to determine the initial state uniquely;
the observer can have access to only part of the Cauchy surface.
In general, it seems a difficult task
to derive the information without fixing the initial state
because $G_\Psi( t^+_1, t^+_2\, ; c )$ depends also on
the initial quantum state $\ket{\Psi\, ; c}$.

In order to distinguish between two boundary conditions $c$ and $c'$
from the behavior of the boundary Wightman functions,
the following inequality should hold for any $\Psi$ and $\Psi'$:
\begin{align}
   G_\Psi( t^+_1, t^+_2\, ; c )
  &= \lim_{r_1,r_2 \rightarrow \infty}~r_1\, r_2~
    \bra{\Psi\,;c}\, \phi(x_1)\, \phi(x_2)\, \ket{\Psi\,;c}
\nonumber \\
  &\ne \lim_{r_1,r_2 \rightarrow \infty}~r_1\, r_2~
    \bra{\Psi'\,;c'}\, \phi(x_1)\, \phi(x_2)\, \ket{\Psi'\,;c'}
  = G_{\Psi'}( t^+_1, t^+_2\, ; c' )~.
\label{inequality}
\end{align}
Otherwise, the boundary Wightman function is {\it degenerate}
for the boundary conditions.
If the function is degenerate,
it is impossible to detect the information
unless we can perfectly control the initial state.

Unfortunately, it is impossible to argue
the behavior of $G_\Psi$ for general initial states $\Psi$.
Thus, we discuss the behavior of $G_\Psi$
for a restricted class of initial states,
which is a class of squeezed states.
The squeezed state $\ket{\bar{0}\, ; c}$ is defined by
$ \bar{a}_n \ket{\bar{0}\, ; c}=0 $ for any $\bar{a}_n$,
where the new annihilation operators $\bar{a}_n$
are given by the Bogoliubov transformation
\begin{align}
  & \bar{a}_n = \sum^\infty_{n'=1} \left(
  \alpha^*_{nn'}\, a_{n'} - \beta^*_{nn'}\, a_{n'}^\dagger \right)~,
\label{Bogo}
\end{align}
which corresponds to the transformation of the mode functions
\begin{align}
  & \bar{f}_n(x)
  = \sum^\infty_{n'=1} \big(~
  \alpha_{nn'}\, f_{n'}(x) + \beta_{nn'}\, f^*_{n'}(x)~\big)~.
\label{eq:Bogo_f}
\end{align}

The Bogoliubov coefficients $\alpha_{nn'}$ and $\beta_{nn'}$ should
satisfy the unitarity condition
\begin{align}
  & \begin{cases}
  {}~\displaystyle{
    \delta_{nn'}
     = \sum^\infty_{m=1} \left( \alpha_{nm}\, \alpha^*_{mn'}
          - \beta_{nm}\, \beta^*_{mn'} \right) }
  \\
  {}~\displaystyle{
    0
     = \sum^\infty_{m=1} \left( \alpha_{nm}\, \beta_{mn'}
          - \beta_{nm}\, \alpha_{mn'} \right) }
  \end{cases}~.
\label{eq:unitarity}
\end{align}
Using this unitarity condition, we can relate the Wightman function
$W_{\bar{0}}$ for the squeezed state to $W_0$ for the vacuum:
\begin{align}
   W_{\bar{0}}( x_1\, ; x_2\, ;c )
  &:= \bra{\bar{0}\,;c}\, \phi(x_1)\, \phi(x_2)\, \ket{\bar{0}\,;c}
\nonumber \\
  &= W_0( x_1\, ; x_2\, ;c ) + 2 \sum_{n,n'=1}^\infty~
    \text{Re}\left[~f_n(x_1)\, \beta^*_{n' n}\, \bar{f}_{n'}(x_2)~
             \right]~.
\label{eq:diff_W-squeeze}
\end{align}
Then, we have%
\footnote{We assume that
the infinite series in Eq.~(\ref{eq:diff_W-squeeze})
is termwise differentiable in $x_1$ and $x_2$, respectively,
because $\beta_{nn'}$ should rapidly decay for large $n$ and $n'$.
}
\begin{align}
  & G_{\bar{0}}( t^+_1, t^+_2\, ;c ) - G_0( t^+_1, t^+_2\, ;c )
  = 4\, h'(t^+_1)\, h'(t^+_2)~\sum_{n,n'=1}^\infty~
    \frac{ 4\, \omega_n\, \omega_{n'} }
         { \sqrt{ ( \pi \omega_n - \sin\pi\omega_n )
           ( \pi \omega_{n'} - \sin\pi\omega_{n'} ) } }
\nonumber \\
  &\times \text{Re}\left[~
    \left( \sum_{m=1}^\infty \beta^*_{m n}\, \beta_{mn'} \right)
      e^{ - i \pi (\omega_n - \omega_{n'})/2 }~
      e^{ 2 i \left\{ \omega_n h(t^+_1)
                    - \omega_{n'} h(t^+_2) \right\} } \right.
\nonumber \\
  &\hspace{2.0cm} \left.
  + \left( \sum_{m=1}^\infty \beta^*_{mn}\, \alpha_{mn'} \right)
      e^{ - i \pi (\omega_n + \omega_{n'})/2 }~
      e^{ 2 i \left\{ \omega_n h(t^+_1)
                    + \omega_{n'} h(t^+_2) \right\} }~\right]~.
\label{eq:diff_G-squeeze}
\end{align}

For simplicity, consider the case where
the Bogoliubov coefficients are diagonal,
$\left\vert\, \alpha_{nn'}\, \right\vert
= \delta_{nn'}\, \cosh\chi_n$ and
$\left\vert\, \beta_{nn'}\, \right\vert
= \delta_{nn'}\, \sinh\chi_n$.
Then, the difference of the boundary Wightman functions
between the squeezed state and the vacuum state is given by
\begin{align}
  & G_{\bar{0}}( t^+_1, t^+_2\, ;c ) - G_0( t^+_1, t^+_2\, ;c )
  = 4\, h'(t^+_1)\, h'(t^+_2)~\sum_{n=1}^\infty~
    \frac{ 2\, \omega_n^2\, \sinh(2 \chi_n) }
         { \pi \omega_n - \sin\pi\omega_n }
\nonumber \\
  & \times \Big( \tanh\chi_n
    \cos\left[~2 \omega_n\, \Delta(t^+_1, t^+_2)~\right]
  + \cos\left[~2 \omega_n \left( H(t^+_1, t^+_2) - \pi/2 \right)
             + \theta_n~\right] \Big)~,
\label{eq:diff_G-diagonal_squeeze}
\end{align}
where $H(s_1, s_2) := h(s_1) + h(s_2)$ and
$\theta_n := \text{arg}(\alpha_{nn}) - \text{arg}(\beta_{nn})$.
On the other hand, from Eq.~(\ref{eq:boundary-Wightman_late}),
we have
\begin{align}
  & G_0( t^+_1, t^+_2\, ; c' ) - G_0( t^+_1, t^+_2\, ; c )
  = - (c' - c)\, \frac{ 2\, h'(t^+_1)\, h'(t^+_2) }{\pi}~
\nonumber \\
  &\times \int_{-\infty}^\infty dy~
    \frac{ e^{ 2 \Delta(t^+_1, t^+_2)\, y} }{ \cosh^2(\pi y/2) }\,
     \frac{y^2}{ \big[ y + c \tanh(\pi y/2) \big]
                 \big[ y + c' \tanh(\pi y/2) \big] }~.
\label{eq:diff-boundary-Wightman_late}
\end{align}
The difference between the Wightman functions for the vacuum states
satisfying different boundary conditions
essentially depends only on
$\Delta(t^+_1, t^+_2) = h(t^+_1) - h(t^+_2)$, not on $H(t^+_1, t^+_2)$.
So, if the Wightman function of the vacuum state characterized by $c$
is represented by the function of a squeezed state
characterized by another parameter $c'$,
the $H(t^+_1, t^+_2)$ dependence
in Eq.~(\ref{eq:diff_G-diagonal_squeeze}) should disappear
for the range $2 \tan^{-1}(r_0)<H(t^+_1, t^+_2)< \pi$.
[Note that $\tan^{-1}(r_0)<h(t^+)<\pi/2$.
See the sentence immediately after Eq.~(\ref{eq:relation-uII}).]

For simplicity, let us examine the $c'=0$ case.
Substituting $\omega_n=2n-1$
into Eq.~(\ref{eq:diff_G-diagonal_squeeze}),
the terms with $H(t^+_1, t^+_2)$ reduce to
\be
f(H):=\frac{8h'(t^+_1)h'(t^+_2)}{\pi}
\sum_{n=1}^\infty (2n-1)\sinh(2\chi_n)
\cos[2(2n-1)(H-\pi/2)+\theta_n].
\ee
This function is anti-periodic with period $\pi/2$,
{\it i.e.}, $f(H+\pi/2)=-f(H)$,
so the function $f$ has to vanish for the range
$2 \tan^{-1}(r_0)<H(t^+_1, t^+_2)< \pi$ for $r_0 < 1$.
This is improbable unless all $\chi_n$ are zero.
Thus, the Wightman function for the vacuum state at nonzero $c$
cannot be represented by the function for any squeezed state at $c=0$.

Unfortunately, it is impossible to show Eq.~(\ref{inequality})
rigorously for general $\Psi$.
But the above argument may suggest that
it is possible to obtain the information about $c$
even when the initial state is unknown.

We have noted thermalization behavior
in Eq.~(\ref{eq:Wightman_remote_late}).
Similar to the discussion there,
the appearance of $h'(t^+_1)$ in Eqs.~(\ref{eq:diff_G-squeeze})
and (\ref{eq:diff-boundary-Wightman_late}) suggests that
the (boundary) Wightman function ``forgets" its initial data
and boundary condition at late time
and it settles into a unique thermal state.
Namely, this factor has the expected form from the Hawking radiation.
The differences left are the minor ones
(for instance, just the overall factor due to the difference
in the boundary condition $c$ and so on.)

\section{Conclusion and discussion}\label{sec:conclusion}

We have investigated how information about the singularity
inside the BTZ black hole formed by gravitational collapse
is restored from the boundary correlator.
A model of timelike singularity is artificially
constructed to represent the information
as one-parameter family of boundary conditions.
When initial quantum state is restricted to a vacuum state,
the parameter $c$ is restored from
the asymptotic behavior of the correlator.
When it is not restricted, it is not clear
whether we can derive the information
since we cannot observe the whole initial data.
If the correlator is {\it degenerate} on the boundary of spacetime,
we cannot derive the information.
In the previous section, we show that the boundary correlator
is not degenerate for a particular class of excited states.
This may imply that one can still derive the information
about singularity inside an AdS black hole
via boundary correlators.
But we are not able to study the issue for generic excited states,
so the result is not conclusive.
%
\begin{figure*}[htb]
  \begin{center}
    \includegraphics[clip]{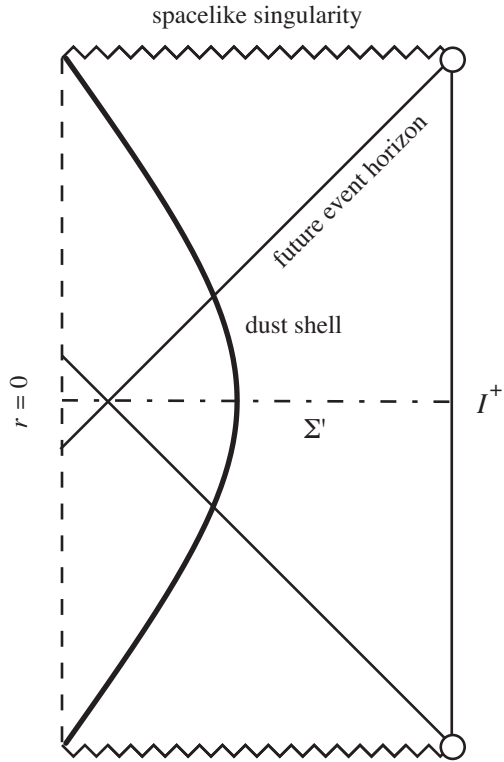}
  \end{center}
\caption{\label{tisy}
Penrose diagram of the time-symmetric spacetime with a
collapsing dust shell}
\end{figure*}%
%


Let us consider the above issue
in the framework of the finite temperature AdS/CFT correspondence.
Originally, the AdS/CFT correspondence is defined
in Euclidean signature, but one may analytically continue it
in Lorentzian signature.
In this case, a vacuum state is naturally selected
by the analytic continuation.
Thus, the Euclidean prescription of AdS/CFT can restrict
initial quantum states in Lorentzian signature.
By such a restriction, it may be possible
to derive the information about singularity inside an AdS black hole
although the prescription does not pick up a unique initial state.
The AdS/CFT correspondence is special in this sense
and probably it is in this sense that
the boundary theory can encode information behind the horizon.

The analytic continuation is also possible
for an AdS black hole formation
if one uses a time-symmetric AdS black hole.
(The analytic continuation must be performed
on the time-symmetric Cauchy surface
as seen in the maximally extended AdS-Schwarzschild spacetime
\cite{Gibbons:1990ns}.)
Figure~\ref{tisy} shows the Penrose diagram of an AdS black hole
formed by the gravitational collapse of a dust shell.
Again one would place the timelike singularity
parametrized by $c$ at $r=0$ in the spacetime.
The parameter $c$ would appear
in the asymptotic behavior of the boundary correlators
and one could extract the information
since the allowed initial quantum state on $\Sigma'$
is restricted.
Note that this timelike singularity
is part of the Euclidean spacetime as well;
From the Euclidean point of view, this is the reason
why the information can be encoded in the boundary correlators.
On the contrary, this spacetime also has a spacelike singularity,
but the spacelike singularity is not part of the Euclidean spacetime.
Thus, extracting the information of the spacelike singularity
is a completely different issue; analyticity seems to play a key role%
\cite{LMR00,KOS02,FHKS,excursion,FHMMRS05}.

Current knowledge of the AdS/CFT correspondence is not enough
to completely resolve the singularity problem.
In particular, we consider the $s$-wave sector
in the BTZ black hole background,
but the ${\rm AdS}_2/{\rm CFT}_1$ correspondence is probably least
understood among the AdS/CFT in various dimensions.
Some aspects are explored, {\it e.g.}, in Refs.~\cite{ads2}.
Most references consider the near-horizon limit of
the extreme four-dimensional Reissner-Nordstr\"{o}m black hole,
and one encounters various difficulties.
We expect similar problems in our case as well
if we seriously try to understand the correspondence
in the ${\rm AdS}_2$ context since our spacetime
is really asymptotically ${\rm AdS}_3$ in disguise.
Probably the Jackiw-Teitelboim black hole is consistently understood
only in the zero mass limit in the ${\rm AdS}_2/{\rm CFT}_1$ context.
One obvious approach to overcome this problem is to go back
to three dimensions
and use the ${\rm AdS}_3/{\rm CFT}_2$ correspondence.
For example, one may compute the correlators
of a three-dimensional scalar field propagating
on the BTZ black hole formed by the gravitational collapse.
In this case, the correlator will involve an infinite sum
due to the infinite number of images, so the expressions
may be more complicated than the one here.

Moreover, the finite temperature AdS/CFT correspondence
is not well-understood quantitatively
compared with zero temperature one.
Precise correspondence between the bulk and the boundary theories
is not known.
We do not pursue this issue in this paper but it is important.
Strong coupling dynamics for finite temperature gauge theories
is in general intractable.
In order to overcome the difficulty,
hydrodynamic description of gauge theory plasmas
using the AdS/CFT correspondence may be useful
since experiments and other tools (such as lattice calculations)
are available there.%
\footnote{See, {\it e.g.}, Ref.~\cite{Kovtun:2004de}
and references therein.
For recent discussion, see, {\it e.g.}, Refs.~\cite{chemical}.}

\begin{acknowledgments}
We would like to thank A. Hosoya for discussions.
The research of M.N.\ was supported in part
by the Grant-in-Aid for Scientific Research (13135224)
from the Ministry of Education, Culture,
Sports, Science and Technology, Japan.
\end{acknowledgments}

\appendix
\section{Evaluation of Eq.~(\ref{eq:def-calW})}\label{sec:appendix}

\begin{figure*}[htb]
  \begin{center}
    \includegraphics[clip]{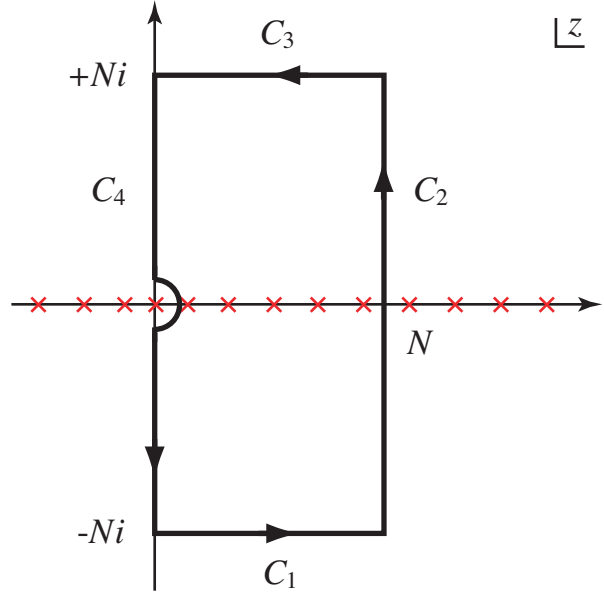}
  \end{center}
\caption{\label{fig2}
The contour of ${\cal W}$ in the complex $z$ plane.
The number $N$ is an arbitrary large positive
number with $N\neq 2n+1$.}
\end{figure*}

In order to evaluate Eq.~(\ref{eq:def-calW}) in the black hole region,
we introduce $p(\omega)$ as
\be
p(\omega):=\omega\cos\frac{\pi \omega}{2}+c\sin\frac{\pi\omega}{2}.
\ee
By using the fact
\be
\frac{1}{\pi \omega_n-\sin\pi\omega_n}=
\frac{c}{2\omega_n\cos(\pi\omega_n/2)}\frac{1}{p'(\omega_n)},
\ee
one can replace the summation of Eq.~(\ref{eq:def-calW})
with the following contour integration in the complex $z$ plane
(Fig.~\ref{fig2})):
\be
\label{contour-eq}
{\cal W}(\alpha)&=&\frac{c}{8\pi i}
\oint_{C_1+C_2+C_3+C_4}\frac{e^{2i\alpha z}}
{z\cos(\pi z/2)p(z)}dz \nonumber \\
&-&\frac{c}{4}\sum_{n=0}^\infty
\mbox{Res}\left[\frac{1}{\cos(\pi z/2)}; z=2n+1 \right]
\frac{e^{2i\alpha z}}{z\, p(z)}\Biggl|_{z=2n+1} \nonumber \\
&=&\frac{c}{8\pi i}
\oint_{C_1+C_2+C_3+C_4}\frac{e^{2i\alpha z}}
{z\cos(\pi z/2)p(z)}dz \nonumber \\
&&+\frac{1}{4\pi}\ln\left|\cot\alpha\right|
+\frac{i}{8}\mbox{sgn}(\alpha).
\ee
%
%
Since $|\alpha|<\pi/2$, $C_1$, $C_2$, and $C_3$ contour integrations
disappear in the large $N$ limit~($N\to \infty$).
The integration along
$C_4$ reduces to
\be
\int_{C_4}\frac{dz}{8\pi i}\frac{e^{2 i\alpha z}}{z\cos(\pi z/2)}
\frac{1}{p(z)}
&=&\text{P} \int^\infty_{-\infty}\frac{dy}{8\pi}\frac{e^{-2\alpha y}}
{y\cosh(\pi y/2)[y\cosh(\pi y/2)+c\sinh(\pi y/2)]}
\nonumber \\
&&-\frac{1}{8}\, \text{Res}\left(
\frac{e^{2 i \alpha z}}{z\cos(\pi z/2)}\,
\frac{1}{p(z)}\right)\Biggl|_{z=0}
\nonumber \\
&=&\frac{1}{2\pi}
\int^\infty_0\frac{dy}{y}\frac{\cosh(2\alpha y)-1}
{y(1+\cosh(\pi y))+c\sinh(\pi y)}-\frac{i\alpha}{2(2+\pi c)},
\nonumber
\ee
where we removed a singularity at $y=0$
by adding an $\alpha$-independent term;
this does not affect the value of $W_0$.


%
\end{document}